\documentstyle[12pt,epsf]{article}

\newcommand{\vs}{\vspace{0.25cm}}

\newtheorem{theorem}{Theorem}
\newtheorem{itlemma}{Lemma}[section]
\newtheorem{itproposition}[itlemma]{Proposition}
\newtheorem{itcorollary}[itlemma]{Corollary}
\newtheorem{itremark}[itlemma]{Remark}
\newtheorem{itremarks}[itlemma]{Remarks}
\newtheorem{itdefinition}[itlemma]{Definition}
\newtheorem{itexample}[itlemma]{Example}

\newenvironment{lemma}{\begin{itlemma}\rm}{\end{itlemma}} 
\newenvironment{remark}{\begin{itremark}\rm}{\end{itremark}} 
\newenvironment{remarks}{\begin{itremarks} \rm}{\end{itremarks}}
\newenvironment{corollary}{\begin{itcorollary}\rm}{\end{itcorollary}}
\newenvironment{proposition}{\begin{itproposition}\rm}{\end{itproposition}}
\newenvironment{definition}{\begin{itdefinition}\rm}{\end{itdefinition}}
\newenvironment{example}{\begin{itexample}\rm}{\end{itexample}}
\newenvironment{fact}{\noindent {\em Fact}. \ \ }{\hfill \medskip}
\newenvironment{proof}{\noindent {\em Proof}.\ \
}{\hspace*{\fill}$\Box$\medskip}
\newenvironment{claim}{\noindent {\em Claim}. \ \ }{\hfill \medskip}
\newcommand{\be}[1]{\begin{equation}\label{#1}}
\newcommand{\ee}{\end{equation}}
\newcommand{\bl}[1]{\begin{lemma}\label{#1}}
\newcommand{\br}[1]{\begin{remark}\label{#1}}
\newcommand{\brs}[1]{\begin{remarks}\label{#1}}
\newcommand{\bt}[1]{\begin{theorem}\label{#1}}
\newcommand{\bd}[1]{\begin{definition}\label{#1}}
\newcommand{\bp}[1]{\begin{proposition}\label{#1}}
\newcommand{\bc}[1]{\begin{corollary}\label{#1}}
\newcommand{\bfact}[1]{\begin{fact}\label{#1}}
\newcommand{\bex}[1]{\begin{example}\label{#1}}
\newcommand{\ec}{\end{corollary}}
\newcommand{\efact}{\end{fact}}
\newcommand{\eex}{\end{example}}
\newcommand{\el}{\end{lemma}}
\newcommand{\er}{\end{remark}}
\newcommand{\ers}{\end{remarks}}
\newcommand{\et}{\end{theorem}}
\newcommand{\ed}{\end{definition}}
\newcommand{\ep}{\end{proposition}}
\newcommand{\epr}{\end{proof}}
\newcommand{\bpr}{\begin{proof}}
\newcommand{\bcl}{\begin{claim}}
\newcommand{\ecl}{\end{claim}}

\newcommand{\bi}{\begin{itemize}}
\newcommand{\ei}{\end{itemize}}
\newcommand{\ben}{\begin{enumerate}}
\newcommand{\een}{\end{enumerate}}
\newcommand{\text}[1]{\hbox{\rm \ #1\ \/}}

\title{Non-stationary Quantum Walks on the Cycle}

\vs

\vs

\author{Domenico
D'Alessandro\thanks{ Department of Mathematics, Iowa State
University, Ames, Iowa, U.S.A.\ \ Electronic address:
daless@iastate.edu}\,, Gianfranco Parlangeli
\thanks { Dipartimento di Ingegneria dell'Innovazione, Universita' del Salento, Lecce, Italy
\ \ Electronic address: gianfranco.parlangeli@unile.it}\, and \,
Francesca Albertini \thanks{ Dipartimento di Matematica Pura ed
Applicata, Universita' di Padova, Italy, \ \ Electronic address:
albertin@math.unipd.it}\,\ \ }

\begin{document}

\maketitle

 \vs

\begin{abstract}

We consider  quantum walks on the cycle in the non-stationary case
where the `coin' operation is allowed to change at each time step.
We characterize, in algebraic terms,  the set of possible state
transfers and prove that, as opposed  to the stationary case, the
associate probability distribution may converge to a uniform
distribution among the nodes of the associated graph.
\end{abstract}

\vs


\vs


\section{Non-stationary quantum walks on the cycle}

Consider a non-oriented graph where all the $N$ nodes have the same
degree $d$ and assume that, at each time step, a walker makes a
choice,  out of a set of $d$ elements, $ \{ 1,...,d \} $, a {\it
(generalized) coin}, with probability $p_1,p_2,..., p_d$,
respectively. The walker starts from a given node of the graph and
moves in a direction determined by  the choice in $\{1,...,d\}$.
After time $t$, the walker will have a probability $P(j,t)$ of being
found in the node $j$, $j=1,...,N$. Such a system is known as a {\it
random walk} on  a graph.
 A {\it quantum walk} is the quantum
counterpart of a random walk in that both the walker and the coin
are seen as quantum systems of dimensions $N$ and $d$, respectively.
At each step an operation $C$ is performed on the coin system and
then an operation is performed on the walker system. The latter
operation depends  on the state of the coin system.

Quantum walks have recently received large attention due to the fact
that they can model quantum algorithms and generate interesting
quantum states. There are several review papers on quantum walks,
their use, dynamics, implementations and generalizations (see, e.g.,
\cite{Kempe1}, \cite{Kendon1}). In most  studies presented so far,
the coin operation $C$ is fixed and repeated at each time step. We
shall call this type of  quantum walks {\it stationary} while
quantum walks where the coin operation is allowed to change at each
time step will be named {\it non-stationary}. Studies exist on how
the parameters  of $C$ affect  the behavior of the quantum walk
\cite{KenCon}. The non-stationary case has been considered
 in both numerical and analytic studies where the coin operation
 is allowed to change at each step according to a prescribed
 sequence or it is random \cite{quattro}, \cite{uno},
 \cite{tre}, \cite{due}. It is shown that, for certain walks,
 the presence of random noise in $C$ at each step, the so called
 {\it unitary noise},  causes a  behavior similar  to the classical
 random walk. A non stationary setting can be considered also for
 other types of random walks such as classical walks on groups (as for example  the
 Heisenberg group)  \cite{French1}, \cite{French2},
 \cite{French3}. These systems  are of current interest as models of quantum
 dynamics. The role of the coin process is played by a dynamical
 system which may be characterized by  a time varying transformation
 therefore giving rise to a non-stationary random walk on a group.
 Similar questions to the ones treated here, in particular
 concerning the set of achievable distributions, can be asked in that
 setting as well.

 In this paper, we approach the study of non-stationary
 quantum walk from the point of view of {\it design} and {\it control} \cite{MikoBook}. We
 consider the coin operation $C$ as a control variable which we can
 change at each step in order to obtain a desired behavior. The
 first questions that arise in this setting are therefore about the
 type of behavior that can be obtained (in particular the type of
 probability distributions) and whether there are significant
 differences with the stationary case. This paper is a first study
 in this direction.

Current proposals for implementations of
 stationary quantum walks (see \cite{Kempe1} and references therein)
   may be modified in order
 to obtain a non-stationary walk. This is discussed for example
 in \cite{tre} for a specific experimental proposal where a variable
 coin operation can be obtained by varying the duration of a laser
 pulse.


\vs

The quantum walk on the cycle is the simplest finite dimensional
quantum walk. The study of stationary quantum walks on the cycle was
started in \cite{Aharonov}. In this case, as  a consequence of the
reversibility of the evolution, the probability distribution
$P(j,t)$ does not converge to a constant  value as $t \rightarrow
\infty$. This is in contrast with the classical random walk on the
cycle whose probability distribution converges to a uniform
distribution independently of the initial state. For this reason,  a
Cesaro type of  alternative probability distribution is introduced
which is the average of $P(j,t)$ over an interval of time $[0,t)$.
With this definition, a uniform limit distribution is obtained for
number of positions $N$ odd, which is independent of the initial
state as long as this one is localized in one given position. For
$N$ even, there is a   much richer behavior and different limit
distributions are obtained for different initial states as discussed
in \cite{cinque}, \cite{sei}. From an experimental point of view,
having a uniform Cesaro type limit probability distribution, means
that there is equal probability of finding the walker in one of the
positions by measuring at a random time over a very large interval.

In dealing with non-stationary walks on the cycle, the first
question concerns the type of (non-Cesaro) probability distributions
that can be obtained. This question is of interest in the use of
random walks for algorithmic purposes. In fact, there exist several
computational algorithms which are based on sampling from a given
set of objects according to a prescribed distribution
\cite{Sinclair}. These algorithms are referred to as {\it randomized
algorithms}. If one uses a quantum walk to implement one of these
algorithms, one can obtain the desired sample by measuring the
position of the walker. One natural question concerns the set of
possible distributions available. We shall answer this question in
Lie algebraic terms in this paper for a non-stationary walk on the
cycle and will show that it is possible that the probability
distribution converges to a {\it uniform} distribution. There are at
least two reasons to consider the uniform distribution with special
attention. One is that it offers an example of a {\it limit}
distribution (in the non-Cesaro sense) which is not available in the
stationary case. In fact we will show that  it is possible to reach
a separable state of the form $|1\rangle \otimes |w\rangle$ where
$|1\rangle$ is a state of a two dimensional coin and $|w\rangle$ is
a state of the walker with all equiprobable positions (cf. formulas
(\ref{tobeshown}) and (\ref{sulmod}) below). Since the coin
operation is arbitrary, we can set it equal to the identity in the
following steps and in the following evolution the walker will just
move from one position to the other in a cycle and the probability
of each position will be uniform. The probability distribution
therefore reaches the uniform value and then stays constant. This is
an example of a behavior different from the stationary case. The
second reason to consider the uniform distribution in more detail is
that this is the limit of the corresponding classical random walk on
the cycle. Therefore, if an algorithm uses this feature of the
classical counterpart, it can be implemented with the non-stationary
quantum walk. For example, a randomized  algorithm which requires
sampling from a uniform distribution can be implemented by measuring
the position of the walker at a large time. If we use the Cesaro
definition of probability distribution we could perform a
measurement but will have to select (again) a random time over a
large interval.



A quantum walk on the cycle is a bipartite quantum system ${\cal C}
\otimes {\cal W}$, where the system ${\cal C}$, the coin, is a two
level system with orthonormal basis states $|+1 \rangle$ and $|-1
\rangle$. The system $\cal W$,
 the walker, is an $N$-level system with orthonormal basis states $|0
\rangle,$ $ |1 \rangle,$...,$|N-1\rangle$. At the $t$-th time-step,
one performs a coin operation of the form $C_t \otimes {\bf 1}$
where $C_t$ is an arbitrary (special) unitary operation on the two
dimensional Hilbert space associated to ${\cal C}$, i.e., an element
of $SU(2)$. This is  followed by a conditional shift $S$ on the
Hilbert space associated to ${\cal W}$ defined as
$$
S|c\rangle \otimes |j \rangle =|c\rangle \otimes |(j +c) \text{mod}
N\rangle.
$$
By considering the standard basis $|e_j\rangle $, $j=1,...,2N$,
defined by $|e_j\rangle:=|1 \rangle \otimes |j-1\rangle$, and
$|e_{j+N}\rangle :=|-1\rangle \otimes |j-1 \rangle$, $j=1,...,N$,
the matrix representation of the operator $C_t \otimes {\bf 1}$ is
$C_t \otimes {\bf 1}_{N \times N}$ where ${\bf 1}_{N \times N}$ is
the $N \times N$ identity,\footnote{We replace this notation by
${\bf 1}$ when there is no ambiguity on the dimension.} $\otimes$
denotes the Kronecker product of matrices, and $C_t \in SU(2)$. The
matrix representation of the operator $S$ is the block diagonal
matrix $\text{diag}(F,F^T)$, where $F$ is the basic circulant
permutation matrix, that is, \be{F} S:=\text{diag}(F,F^T), \qquad
F:=\pmatrix{0 & 0 & \cdot & \cdot & \cdot & 0 & 1 \cr 1 & 0 & \cdot
& \cdot & \cdot & 0 & 0 \cr 0 & 1 & \cdot & \cdot & \cdot & 0 & 0
\cr \cdot  & \cdot & \cdot & \cdot & \cdot & \cdot & \cdot \cr \cdot
& \cdot & \cdot & \cdot & \cdot & \cdot  & \cdot \cr \cdot & \cdot &
\cdot & \cdot & \cdot & \cdot  & \cdot \cr 0 & \cdot  & \cdot &
\cdot & \cdot & 1 & 0}. \ee
%

\vs

The probability of finding the walker in state $|j-1\rangle$,
$j=1,...,N$, is the sum of the probabilities of finding the state of
the composite system ${\cal C} \otimes {\cal W}$ in $|1\rangle
\otimes |j -1\rangle:=|1,j-1\rangle$ and  $|-1\rangle \otimes |j
-1\rangle:=|-1,j-1\rangle$. That is, if $|\psi\rangle$ is the state
of the composite system, \be{proba} P(j-1,t)=|\langle \psi(t)| 1 ,j
-1\rangle|^2+|\langle \psi(t)| -1 , j -1\rangle|^2=|\langle
\psi(t)|e_j\rangle|^2 + |\langle \psi(t)|e_{j+N}\rangle|^2. \ee By
writing \be{psiw} |\psi \rangle:=\sum_{k=1}^{2N} \alpha_k
|e_k\rangle, \qquad \sum_{k=1}^{2N} |\alpha_k|^2=1,  \ee we have
\be{prt} P(j-1,t)=|\alpha_j|^2+ |\alpha_{j+N}|^2, \qquad j=1,...,N.
\ee \vs In what follows we shall make the following standing
assumption. \vs

\noindent {\bf Assumption:} N is an {\it odd} number.

\vs

We use this technical assumption in several steps and in particular
in Theorem \ref{evolutions} to show that the matrix $S$ defined in
(\ref{F}) is in a certain Lie group (cf. formula (\ref{TogetS})). In
the stationary case, problems with the $N$ even case arise from the
degeneracy of eigenvalues in the basic $S (C \otimes {\bf 1})$
operation. As we have mentioned, this leads to a very different and
richer behavior with respect the $N$ odd case.

\section{Characterization of the admissible evolutions}

In this section, we characterize the set of unitary evolutions
available for a non-stationary quantum walk on the cycle, that is,
the set of available state transfers.  This is the set of finite
products of operators of the form $S(C \otimes {\bf 1}_{N \times
N})$ where $S$ is defined in (\ref{F}) and $C \in SU(2)$. We denote
such a set by ${\bf G}$. The set ${\bf G}$ is a group. It is in fact
a Lie group as it is shown in the following theorem. In order to
state this theorem, we need to recall some properties of circulant
matrices \cite{Davis} and define two Lie algebras.

Circulant $N \times N$ matrices with complex entries form a vector
space over the real numbers. Each matrix is determined by the first
row since  all the other rows can be obtained by cyclic permutation
of the first one. Moreover every complex circulant matrix $R$ can be
written as linear combination with complex coefficients of the basic
permutation matrix $F$ defined in (\ref{F}) and its powers from $0$
to $N-1$, i.e., \be{Expa} R:=\sum_{l=0}^{N-1}a_l F^l,   \ee with $N$
complex coefficients $a_0,...,a_{N-1}$. All the circulant matrices
commute. If we require that $R$ is not only circulant but also
skew-Hermitian then we must have \be{skewHer} R^\dagger=a_0^* {\bf
1}+\sum_{l=1}^{N-1}a_l^*F^{lT}=-R=-a_0{\bf 1}-
\sum_{l=1}^{N-1}a_lF^l, \ee and with a change of index $l
\rightarrow N-l$ and using ${F^{N-l}}^T=F^l$, we have \be{skewHer2}
R^\dagger=a_0^* {\bf 1}+\sum_{l=1}^{N-1}a_{N-l}^*F^l=-a_0{\bf 1}-
\sum_{l=1}^{N-1}a_lF^l. \ee This gives the relations \be{rt}
a_0^*=-a_0, \qquad a_{N-l}^*=-a_l, \quad l=1,...,\frac{N-1}{2}. \ee
Equations (\ref{skewHer2}) constitute $N$ independent relations on
the $2N$ real parameters of $R$ and show that the space of
skew-Hermitian circulant matrices is a real vector space of
dimension $N$.

\vs

Denote by $\cal L$ the Lie algebra spanned by the $2N \times 2N$
skew-Hermitian matrices of the form \be{ghi} L_1:=\pmatrix{R & 0 \cr
0 & -R} \quad \text{and} \quad  L_2:=\pmatrix{0 & Q \cr -Q^\dagger &
0}, \ee with $R$ a skew-Hermitian circulant $N \times N$ matrix and
$Q$ a general circulant matrix. It is easily seen that this is in
fact a Lie algebra of (real) dimension $3N$; the fact that it is
closed under Lie bracket being a consequence of the fact that the
product of two circulant matrices is another circulant matrix.
Notice, in particular, that matrices of the type $L_1$ form an
Abelian subalgebra of dimension $N$. We denote by $e^{\cal L}$ the
connected Lie group associated to ${\cal L}$.

\vs

\bt{evolutions} The set ${\bf G}$ of possible evolutions of a
non-stationary quantum walk is the Lie group $e^{\cal L}$. \et

\bpr We define an auxiliary Lie algebra ${\cal L}'$, prove that
${\bf G}=e^{{\cal L}'}$ and then prove that ${\cal L}={\cal L}'$.
The claim then follows from the correspondence between Lie groups
and Lie algebras. We denote by ${\cal L}'$ the Lie algebra generated
by the set \be{EFFE1} {\cal F}:=\{ su(2)\otimes {\bf 1}, S \left(
su(2) \otimes {\bf 1} \right) S^T,..., S^{N-1}\left(su(2) \otimes
{\bf 1} \right) S^{(N-1)T}\}, \ee where $S$ is defined in (\ref{F})
and $T$ denotes transposition.

To show that ${\bf G} \subseteq e^{{\cal L}'}$, it is enough to show
that
 both $C \otimes {\bf 1}$, $C \in SU(2)$,  and $S$ are in $e^{{\cal L}'}$. This fact is obvious
 for $C \otimes {\bf 1}$, since this is the exponential of an
 element in $su(2) \otimes {\bf 1}$. For $S$, we consider the
 elements $\pmatrix{0 & -1 \cr 1 &0} \otimes {\bf 1}$ and
 $S^{\frac{N-1}{2}}\left(\pmatrix{0 & 1 \cr -1 &0} \otimes {\bf
 1}\right)
 S^{\left(\frac{N-1}{2}\right)T}$, both in $e^{{\cal L}'}$, and calculate with
 (\ref{F})
 \be{TogetS}
\left[\pmatrix{0 & -1 \cr 1 &0} \otimes {\bf 1}
\right]\left[S^{\frac{N-1}{2}}\left(\pmatrix{0 & 1 \cr -1 &0}
\otimes {\bf 1}\right)
 S^{\left(\frac{N-1}{2}\right)T}\right]=
 \ee
 $$\pmatrix{F^{(N-1)T} & 0 \cr 0 & F^{N-1}}=\pmatrix{F & 0 \cr 0 &
 F^T}:=S.$$ We have used $F^{(N-1)T}=F$.

 To show that $e^{{\cal L}'} \subseteq {\bf G}$
 it is enough to show that every element of the type $S^j \left(
X\otimes {\bf 1} \right)S^{jT}$, with $X \in SU(2)$, $j=0,...,N-1$,
can be written as the finite product of elements of the form
$S\left(C \otimes {\bf 1}\right)$ with $C \in
SU(2)$.\footnote{Recall that every element of a connected Lie group
can be obtained as the finite product of exponentials of a set of
generators of the corresponding Lie algebra (see, e.g., \cite{JS})
and the exponential map is surjective on $SU(2)$ (see, e.g.,
\cite{SagleWalde}).} This is readily seen because, with $X\in
SU(2)$, for every $j$,
 \be{obvrel}
S^j \left(X \otimes {\bf 1}_{N \times N} \right) S^{jT}= \left( S^j
\left( X \otimes {\bf 1}_{N \times N}\right) \right) \left( S^{N-j}
\left({\bf 1}_{2 \times 2} \otimes {\bf 1}_{N \times N} \right)
\right). \ee

\vs

To conclude the proof, we show that ${\cal L}={\cal L}'$ showing
that ${\cal F} \subseteq {\cal L}$ and a basis of ${\cal L}$ can be
obtained as (repeated) Lie brackets and-or linear combinations of
elements of ${\cal F}$ in (\ref{EFFE1}). A general matrix in ${\cal
F}$ has the form, with $A \in su(2)$, \be{hjk} S^j \left( A \otimes
{\bf 1}_{N \times N} \right) S^{jT}= \ee $$\pmatrix{F^j & 0 \cr 0 &
F^{jT}} \pmatrix{ib{\bf 1}_{N \times N}& \alpha {\bf 1}_{N \times N}
\cr -\alpha^* {\bf 1}_{N \times N} & -ib{\bf 1}_{N \times N}}
\pmatrix{F^{jT} & 0 \cr 0 & F^{j}} = \pmatrix{ib {\bf 1}_{N \times
N} & \alpha F^{2j} \cr -\alpha^* (F^{2jT}) & -ib {\bf 1}_{N \times
N}}, $$
 with general $b$ real and $\alpha $ complex, $j=0,...,N-1$.  This is
clearly in ${\cal L}$. Elements of the form $L_2$ in (\ref{ghi}) are
real linear combinations of elements of the form $\pmatrix{0 &
\gamma F^{k} \cr -\gamma^* F^{kT} & 0}$ which are of the form in
(\ref{hjk}) with $b=0$, $\gamma=\alpha$ and $j=\frac{k}{2}$ for $k$
even and $j=\frac{N+k}{2}$ for $k$ odd. A basis for the real
elements of the type $L_1$ is given by the $\frac{N-1}{2}$ linearly
independent elements \be{lop} \pmatrix{F^j-F^{jT} & 0 \cr 0 &
-(F^j-F^{jT})}, \qquad j=1,...,\frac{N-1}{2}.   \ee These are
obtained as Lie brackets of $\pmatrix{0&{\bf 1}_{N \times N} \cr
-{\bf 1}_{N \times N}&0}$ and $\pmatrix{0&F^j \cr -F^{jT}&0}$ which
are both of type $L_2$. A basis for the purely imaginary elements of
type $L_1$ is given by the $\frac{N+1}{2}$ linearly independent
elements of the type \be{ppp} \pmatrix{i(F^j+F^{jT}) & 0 \cr 0 & -i
(F^j+F^{jT})}, \qquad j=0,...,\frac{N-1}{2}, \ee which are obtained
as Lie brackets of $\pmatrix{0 & F^j \cr -F^{jT} &0}$ and
$\pmatrix{0 & i {\bf 1}_{N \times N} \cr i {\bf 1}_{N \times N} &
0}$. This completes the proof of the theorem. \epr

\section{Obtaining the uniform distribution}

The Lie group ${\bf G}= e^{\cal L}$, having dimension $3N$, is not
isomorphic, for $N \geq 3$, to $SU(2N)$ (which has dimension
$4N^2-1$) nor to $Sp(N)$ (which has dimension $N(2N+1)$). Therefore
${\bf G}=e^{\cal L}$ is not transitive on the complex sphere of
dimension $2N$ which means that there are state transfers for the
quantum system of coin and walker which are not induced by any
transformation in ${\bf G}$ \cite{confracon}. Some state transfers
are of special interest. In particular, we are interested in whether
a state of the form \be{iopl} |\psi_{in}\rangle:=|\psi_{coin}
\rangle \otimes |0 \rangle, \ee that is, a state corresponding to
the walker with certainty in position $|0\rangle$, can be
transferred to a state corresponding to the uniform distribution.
This is a state where the probability $P(j-1,t)$ in (\ref{proba}) is
equal to $\frac{1}{N}$, for every $j=1,...,N$, at some $t$, that is,
the walker is found in any position with the same probability.
Since, $\forall \, C \in SU(2)$, $C \otimes {\bf 1}_{N \times N} \in
e^{\cal L}$, we can assume, without loss of generality, that
$|\psi_{coin}\rangle $ in (\ref{iopl}) is $|1\rangle$ so that the
problem is to transfer the state $|e_1 \rangle:=[1,0,...,0]^T$ to a
state with the desired property. We shall show in the following that
such a state transfer is possible.

\bt{Transfer} There exists a matrix $L$ in ${\cal L}$ such that
\be{tobeshown} e^L|e_1 \rangle=\pmatrix{r_1 \cr r_2 \cr \vdots \cr
r_N \cr 0 \cr 0 \cr \vdots \cr 0}, \ee \et where \be{sulmod}
|r_1|^2=|r_2|^2=\cdot \cdot \cdot =|r_N|^2=\frac{1}{N}. \ee

In order to prove this theorem we first prove a lemma. Recall the
definition of the Fourier matrix $\Phi$ of order $N$ (see, e.g.,
\cite{Davis}). This is defined so  that its conjugate transposed is
\be{fi} \Phi^\dagger:=\frac{1}{\sqrt{N}}\pmatrix{ 1 & 1 & 1 & 1&
\ldots & 1 \cr 1 & \omega & \omega^ 2 & \omega^3 & \ldots &
\omega^{N-1} \cr 1 & \omega^2 & \omega^4 & \omega^6 & \ldots &
\omega^{2(N-1)} \cr 1 & \omega^3 & \omega^6 & \omega^9 & \ldots &
\omega^{3(N-1)} \cr \vdots & \vdots & \vdots & \vdots &  & \vdots
\cr 1 & \omega^{N-1} & \omega^{2(N-1)} & \omega^{3(N-1)} & \ldots &
\omega^{(N-1)(N-1)}}, \ee where $\omega$ is the $N$-th root of the
unity, that is $\omega:=e^{i \frac{2 \pi}{N}}$. The Fourier matrix
$\Phi$ is unitary.

\bl{basiclemma} Define \be{defp} x_l:=\frac{l(l-1)\pi}{N}, \qquad
l=0,1,...,N-1.  \ee \el Then \be{tyh} \pmatrix{r_1 \cr r_2 \cr r_3
\cr \vdots \cr r_N}:=\frac{1}{\sqrt{N}}\Phi^\dagger \pmatrix{e^{i
x_0} \cr e^{ix_1} \cr e^{ix_2} \cr \vdots \cr e^{ix_{N-1}}}\ee has
the property (\ref{sulmod}).

\bpr {}From (\ref{fi}) and (\ref{tyh}), we obtain \be{erreh}
r_h=\frac{1}{N} \left(1+\sum_{l=1}^{N-1} \omega^{({h-1})l} e^{ix_l}
\right), \qquad h=1,...,N.  \ee This, using the definition of
$\omega$, gives \be{err} r_h=\frac{1}{N} \left(1+\sum_{l=1}^{N-1}
e^{\frac{i2\pi({h-1})l}{N}} e^{ix_l} \right). \ee We calculate
$|r_h|^2$, $h=1,...,N$,  as \be{normaquad}
|r_h|^2=r_h^*r_h=\frac{1}{N^2}\sum_{l_1 , l_2=0}^{N-1} e^{i \frac{2
\pi}{N}(l_2 -l_1)(h-1)} e^{i(x_{l_2}-  x_{l_1})} =\ee
$$
\frac{1}{N}+\frac{2}{N^2}\sum_{\{l_1, l_2\}=0}^{N-1} \text{Re}
\left( e^{i\frac{2 \pi}{N}(l_2-l_1)(h-1)} e^{i(x_{l_2}-x_{l_1})}
\right).
$$
The sum in the last term is intended over all the pairs  of indices
$\{ l_1,l_2 \}$, with $l_1 \not= l_2$,  where only one is chosen
between $\{l_1, l_2\}$ and $\{l_2, l_1\}$. Because of the presence
of the real part `$\text{Re} $' it is not important which pair is
chosen. We now show that, with the choice (\ref{defp}), the last
term of this expression is zero for every $h$, which will prove the
claim that $|r_h|^2=\frac{1}{N}$.

It is convenient to re-write the sum by regrouping elements
corresponding to $l_2-l_1=p \text{mod} N$, for $p=1,...,N-1$.  This
means $l_2-l_1=p$ or $l_1-l_2=N-p$. We have \be{hjl} \sum_{\{l_1,
l_2\}=0}^{N-1} \text{Re} \left( e^{i\frac{2 \pi}{N}(l_2-l_1)(h-1)}
e^{i(x_{l_2}-x_{l_1})} \right)=\ee
$$\sum_{p=1}^{N-1} \text{Re} \left( \sum_{l_2-l_1=p}
e^{i(l_2-l_1)(h-1) \frac{2\pi}{N}} e^{i(x_{l_2}-x_{l_1})}+
\sum_{l_1-l_2=N-p} e^{i(l_2-l_1)(h-1) \frac{2\pi}{N}}
e^{i(x_{l_2}-x_{l_1})}\right). $$ Doing the substitution $l_1=l$ and
$l_2=l+p$ in the first term of the sum and the substitution $l_1=l$
and $l_2=l-(N-p)$ in the second term, this sum becomes \be{somme}
\sum_{p=1}^{N-1} \text{Re} \left( e^{ip(h-1)\frac{2\pi}{N}} \left(
\sum_{l=0}^{N-1-p} e^{i(x_{l+p}-x_l)}+ \sum_{l=N-p}^{N-1}
e^{i(x_{l-(N-p)}-x_l)} \right)\right). \ee We now show that, with
the choice (\ref{defp}), the content of the innermost parenthesis in
the above expression, i.e.,  \be{Q} M:=M(p):=\sum_{l=0}^{N-1-p}
e^{i(x_{l+p}-x_l)}+ \sum_{l=N-p}^{N-1} e^{i(x_{l-(N-p)}-x_l)}, \ee
is zero for each $p$ which will conclude the proof of the Lemma.
Replacing (\ref{defp}) in (\ref{Q}) and after some algebraic
manipulations, we obtain \be{QE} M(p)=\sum_{l=0}^{N-1-p} e^{i\frac{2
\pi}{N}\left(\frac{p(p-1)}{2}+pl \right)}+
\sum_{l=N-p}^{N-1}e^{i\frac{2
\pi}{N}\left(\frac{(N-p)(N-p+1)}{2}-(N-p)l \right)}=
\sum_{l=0}^{N-1} e^{i\frac{2 \pi}{N}\left(\frac{p(p-1)}{2}+pl
\right)}. \ee Thus we have \be{almostfin} M(p)=e^{i \frac{2 \pi}{N}
\frac{p(p-1)}{2}} \sum_{l=0}^{N-1} e^{i \frac{2\pi pl}{N}} = e^{i
\frac{2 \pi}{N} \frac{p(p-1)}{2}}\frac{1-e^{i 2 \pi p}}{1-e^{i\frac{
2 \pi p}{N}}}=0 \  \ \ \ \forall p \,  \neq 0 \text{mod} N. \ee This
concludes the proof of the Lemma. \epr

We are now ready to prove Theorem  \ref{Transfer}

\bpr (Proof of Theorem \ref{Transfer}) We choose $L$ as a matrix of
the form $L_1$ in (\ref{ghi}) so that $e^{L}$ has the form
\be{formeL} e^{L}=\pmatrix{e^{R} & 0 \cr 0 & e^{-R}},  \ee with $R$
a general skew-Hermitian $N \times N$ circulant matrix. The problem
is therefore to find a circulant matrix $R$ so that \be{pa}
e^R\pmatrix{1 \cr 0 \cr \vdots \cr 0}= \pmatrix{r_1 \cr r_2 \cr
\vdots \cr r_N}, \ee
 with $r_1,...,r_N$ satisfying (\ref{sulmod}). Any circulant matrix
 $R$ is diagonalized by the Fourier matrix (\ref{fi}) of the
 corresponding dimension,  that
 is,
 \be{pppa}
R=\Phi^\dagger \Lambda \Phi, \ee with $\Lambda$ diagonal. Conversely
every matrix of the form on the right hand side is circulant
\cite{Davis}. Moreover if  $\Lambda=\text{diag} (i
\lambda_0,i\lambda_1,...,i \lambda_{N-1})$, with $\lambda_l$,
$l=0,...,N-1$ real numbers, $R$ is skew-Hermitian. In this case, we
have \be{hjh4} e^{R} \pmatrix{1 \cr 0 \cr \vdots \cr 0}=\Phi^\dagger
e^{\Lambda} \Phi \pmatrix{1 \cr 0 \cr \vdots \cr 0}=\Phi^\dagger
e^{\Lambda}\frac{1}{\sqrt{N}}\pmatrix{1 \cr 1 \cr \vdots \cr
1}=\Phi^\dagger \frac{1}{\sqrt{N}} \pmatrix{e^{i \lambda_0} \cr e^{i
\lambda_1}\cr \vdots \cr e^{i \lambda_{N-1}}}. \ee Choosing
$\lambda_l=x_l$, $l=0,...,N-1$ with the definition (\ref{defp}), the
theorem follows from Lemma \ref{basiclemma}.  \epr

Other  states with the same property can be
 obtained by  applying a transformation
 $U \otimes {\bf 1}$, $U \in SU(2)$, which is in ${\bf G}$. In
 particular notice that the state (\ref{tobeshown}) is a separable state.

\section{Conclusion}

Non-stationary quantum walks have properties which distinguish them
from stationary ones. Moreover they are amenable of study with the
methods of quantum control. In fact, several problems,  such as
obtaining a given evolution, can be seen as control problems where
the evolution of the coin plays the role of the control. In this
paper we have shown that, opposite  to the stationary case,  a
non-stationary  quantum walk on the cycle may converge to a constant
distribution and in particular to a uniform distribution as for
classical random walks. A constructive approach to achieve this and
other evolutions of interest for general quantum walks will be the
subject of future research.

\vs

\noindent {\bf Acknowledgement} D.D. is grateful to Mark Hillery for
helpful conversations on the topic of quantum walks. The authors
also thank the reviewers for useful suggestions.  D. D'Alessandro's
research was supported by NSF under Career Grant ECS-0237925.

\vs

\end{document}